\documentclass[a4paper,fleqn]{cas-dc}

\usepackage[numbers]{natbib}

\usepackage[figuresright]{rotating}
\usepackage{float}
\usepackage{amsmath,amssymb,amsfonts}
\usepackage{algorithmic}
\usepackage{graphicx}
\usepackage{textcomp}

\usepackage{listings}
\usepackage{color}
\usepackage{xspace}
\usepackage{xcolor}
\usepackage{float}
\usepackage{booktabs}
\usepackage{caption}
\DeclareCaptionLabelSeparator{custom}{.}
\usepackage{cleveref}
\crefname{figure}{\textcolor{cyan}{Fig.}}{\textcolor{cyan}{Fig.}}
\crefname{table}{\textcolor{cyan}{Table}}{\textcolor{cyan}{Table}}
\crefname{section}{\textcolor{cyan}{Section}}{\textcolor{cyan}{Section}}
\usepackage{stfloats}
\usepackage{url} 
\usepackage{framed} 
\usepackage[T1]{fontenc} 
\usepackage{multirow}
\usepackage{multicol}
\usepackage[ruled,linesnumbered,noend]{algorithm2e}
\usepackage{threeparttable}
\usepackage{subfigure}
\usepackage{mathrsfs}
\usepackage{inconsolata}
\usepackage{pifont}
\usepackage{balance}
\usepackage{verbatim}
\usepackage{color, colortbl}
\usepackage{tikz}
\usepackage{makecell}
\usepackage{hyperref}
\usepackage{indentfirst}
\usepackage{ragged2e}

\newcommand{\approach}{{JIT-BiCC}\xspace} 
\newcommand{\model}{{BiCC-BERT}\xspace} 

\newcommand{\new}[1]{{#1}}

\begin{document}
\begin{sloppypar}

\let\WriteBookmarks\relax
\def\floatpagepagefraction{1}
\def\textpagefraction{.001}

\shorttitle{Just-In-Time Software Defect Prediction via Bi-modal Change Representation Learning}

\shortauthors{Yuze Jiang et~al.}

\title [mode = title]{Just-In-Time Software Defect Prediction via Bi-modal Change Representation Learning} 

\author[1,2]{Yuze Jiang}
\ead{jiang.2896@osu.edu}

\author[1]{Beijun Shen}
\ead{bjshen@sjtu.edu.cn}

\author[1]{Xiaodong Gu\corref{cor1}}
\ead{xiaodong.gu@sjtu.edu.cn}
\cortext[cor1]{Corresponding Author: Xiaodong Gu}

\address[1]{School of Electronic Information and Electrical Engineering, Shanghai Jiao Tong University, Shanghai, China}

\address[2]{Department of Computer Science and Engineering, The Ohio State University, Columbus, OH, United States}


\begin{abstract}
\noindent For predicting software defects at an early stage, researchers have proposed just-in-time defect prediction (JIT-DP) to identify potential defects in code commits. The prevailing approaches train models to represent code changes in history commits and utilize the learned representations to predict the presence of defects in the latest commit. However, existing models merely learn editions in source code, without considering the natural language intentions behind the changes. This limitation hinders their ability to capture deeper semantics.
To address this, we introduce a novel bi-modal change pre-training model called \model. \model is pre-trained on a code change corpus to learn bi-modal semantic representations. To incorporate commit messages from the corpus, we design a novel pre-training objective called Replaced Message Identification (RMI), which learns the semantic association between commit messages and code changes. Subsequently, we integrate \model into JIT-DP and propose a new defect prediction approach --- \approach. By leveraging the bi-modal representations from \model, \approach captures more profound change semantics.
We train \approach using 27,391 code changes and compare its performance with 8 state-of-the-art JIT-DP approaches. The results demonstrate that \approach outperforms all baselines, achieving a 10.8\% improvement in F1-score. This highlights its effectiveness in learning the bi-modal semantics for JIT-DP.

\end{abstract}


\begin{highlights}

\item Construct \model, a novel code change pre-training model that extracts bi-modal semantic information on code changes.

\item Design a novel pre-training objective called Replaced Message Identification (RMI), allowing the model to explicitly learn the semantic association between commit messages and code changes.

\item Propose \approach, an approach for just-in-time defect prediction (JIT-DP) based on semantic representations extracted by the pre-trained \model.

\item Significantly outperform the state-of-the-art approaches on the JIT-Defects4J dataset.
\end{highlights}


\maketitle

\section{Introduction}

Software defect prediction (SDP), namely, identifying potential defects before software release, has been an indispensable technique in software development~\cite{first_who_propose_SDP}. SDP helps developers predict potential defects in the early stages of the software development cycle so that they can take measures to mitigate the \new{defects~\cite{SDP_practical_value}}. The state-of-the-art \new{approaches~\cite{GirayBKBT23, HeSC16}} mainly employ machine learning techniques to estimate the probability or number of bugs that a given source code file contains. 

\begin{figure}[htbp]
\centerline{\includegraphics[width=\linewidth]{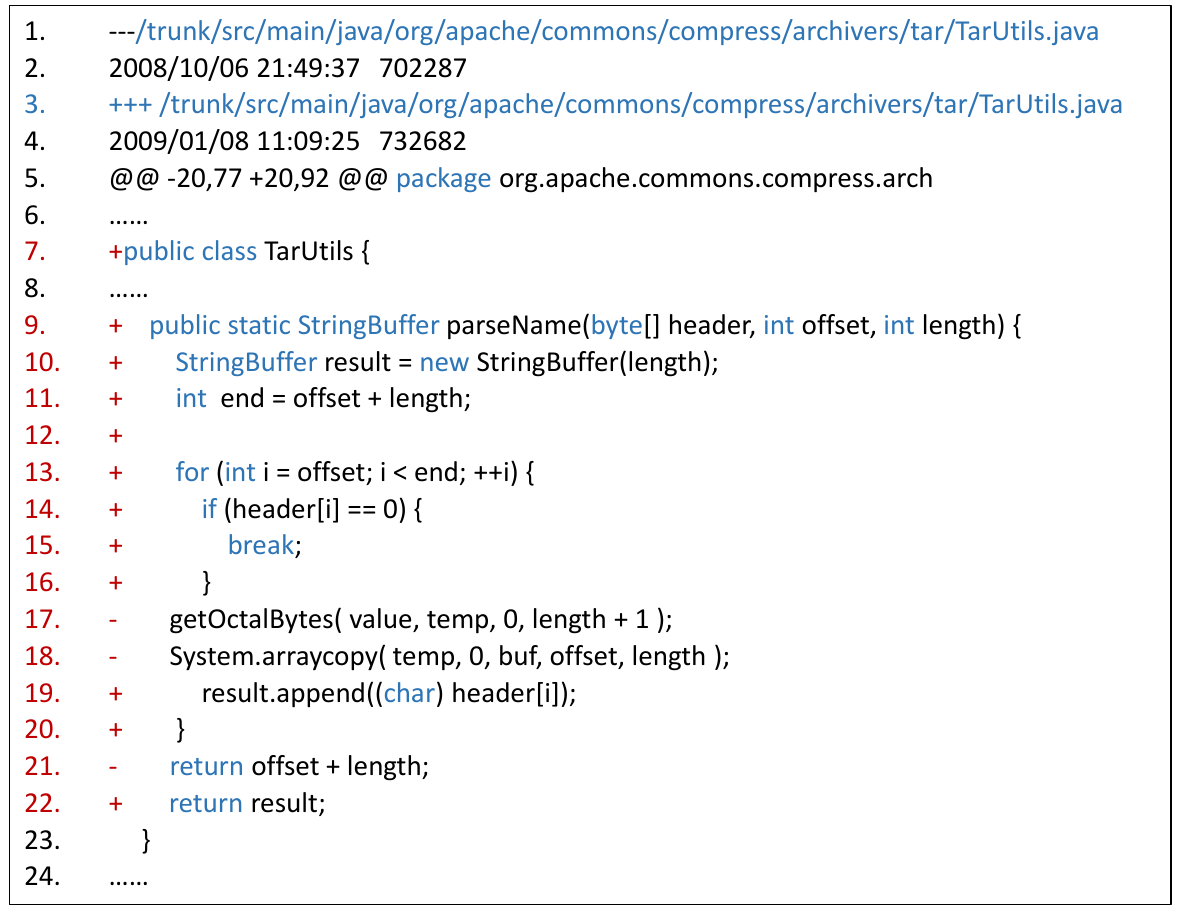}} 
\caption{An illustration of a defect-inducing code change}
\label{example_defective_commit_1}
\end{figure}

\begin{figure}[htbp]
\centerline{\includegraphics[width=\linewidth]{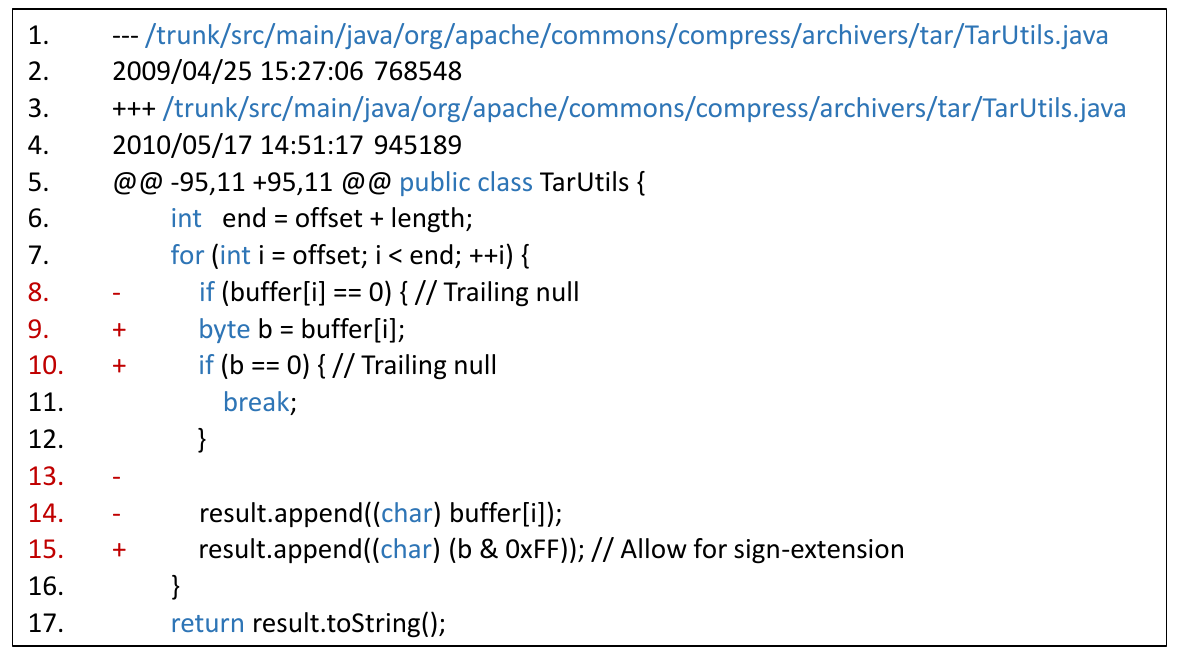}} 
\caption{Fixing patch to the defective code change in Figure 1}
\label{example_defective_commit_2}
\end{figure}

Recently, just-in-time defect prediction (JIT-DP), which aims to predict defects immediately after code changes, has emerged as a major SDP technique~\cite{JIT-DP_survey}. Source code is constantly changing in the software development process, which inevitably introduces software defects~\cite{change_induce_bug, 2014Impact}. It is desirable to predict potential bugs right after developers commit the code to version control systems (e.g., Git) so that developers can fix them \new{promptly~\cite{Fix_promptly}}. Figures \ref{example_defective_commit_1} and \ref{example_defective_commit_2} provide examples illustrating how code changes could introduce software defects, specifically, the defective commit \#732682 and its fixing commit \#945189 in Apache's project commons-compress.


In commit \#732682 on 2009/01/08, developers created a new method named "parseName" with an added line "result.append((char) header[i]);" (Line 19) to store the header's entry name for the returning value. However, it was not until one year later that they discovered that this line would incur a name round-trip problem. Subsequently, they fixed this issue in commit \#945189 by modifying this line to "result.append((char) (b \& 0xFF));" (Line 15) to allow for sign-extension.

JIT-DP is usually conceptualized as a binary classification \new{problem~\cite{JITDP_definition}}: given a code change represented as features, a machine-learning classifier is employed to determine whether it introduces a software defect or not. Thus, the main objective for JIT-DP is to learn code change representations, namely, representing code change as feature vectors. There have been numerous features for change representations~\cite{jitfine}, including expert features (also known as metric features) and semantic features. 

Despite showing promising potential, representing code changes is notably challenging. First, current change representation techniques, such as CC2Vec~\cite{cc2vec} and JIT-Fine~\cite{jitfine}, often rely on only a limited amount of manually labeled data samples to train their defect prediction models. This is usually insufficient for the models to acquire the accurate semantics in code changes with complex patterns and associations.

Moreover, contemporary \new{approaches~\cite{jitfine,cc2vec,CodeReviewer} }solely concern the unimodal features in source code. This restricts their capability to capture the semantic associations between commit messages and code changes.
\new{Previous work~\cite{how_far_are_we} has} demonstrated that commit messages can provide crucial information for understanding code changes, through the gradient-weighted class activation mapping (Grad-CAM) algorithm~\cite{Grad-cam}. However, existing JIT-DP approaches either completely omit commit messages in the defect prediction model~\cite{how_far_are_we} or oversimplify the processing of commit messages~\cite{jitfine, deepjit}. Therefore, they fail to leverage the explicit semantic associations between commit messages and code changes in change representations.

To alleviate these limitations, in this paper, we propose \model, a novel bi-modal change representation model. To enhance the semantic representation of code changes, we design a novel pre-training objective called Replaced Message Identification (RMI), which replaces the commit message and asks the model to predict the replacement. This forces the model to align the semantics between commit messages and code changes.

Based on \model, we present \approach, an approach for JIT-DP by integrating \model into the conventional defect prediction framework. Our approach takes as input code changes as well as their corresponding commit messages and extracts bi-modal semantic features using \model. The semantic features are fused with traditional expert features through fully-connected neural networks. Finally, the two sources of features are concatenated and fed to a classifier for defect prediction.

We evaluate \approach on the JIT-Defects4J dataset which consists of 21 open-source projects with 27,391 changes. We measure the performance using the F1-score and AUC. Experimental results demonstrate that \approach significantly outperforms baseline approaches, with over 10.8\% improvement in terms of the F1-score.

The main contributions of this paper can be summarized as follows:

\begin{itemize}
\item[1)] We propose a novel bi-modal change representation model in which we design a novel pre-training objective called Replaced Message Identification (RMI). The new pre-training objective enables explicit learning of the association between commit messages and code changes.

\item[2)] We propose \approach, a novel approach for JIT-DP by incorporating \model with a defect prediction framework. Our approach captures more profound change semantics by learning representations from both code change and their corresponding commit messages. 

\item[3)] We extensively evaluate \approach on the JIT-Defects4J dataset using F1-score and AUC. Experimental results demonstrate that \approach significantly outperforms state-of-the-art approaches.
\end{itemize}

The rest of this paper is organized as follows. 
Section 2 presents the background related to our work as well as the motivation of \model and \approach. Section 3 describes our proposed model \model and approach \approach in detail. Section 4 presents our experimental setting including compared baselines and considered performance measures. Section 5 reports the results of our experiment. Section 6 introduces the related works of representation learning and JIT-DP. Section 7 concludes this paper and points out potential future works. 

\section{Background and Motivation}

\subsection{Just-In-Time Defect Prediction}
Just-in-time defect prediction (JIT-DP) aims at predicting whether a code commit contains \new{defects~\cite{JITDP_definition}}. JIT-DP assumes that historical code changes that introduce defects have similarities with future changes, and trains a machine learning model to identify defect-prone code \new{changes~\cite{Historical_change}}. 

Formally, for a dataset $D=\{(x_1, y_1),...,(x_N,y_N)\}$ of $N$ code commits, where $x_i$ is a code commit and $y_i\in\{0,1\}$ represents whether $x_i$ is defective (=0) or not (=1), JIT-DP aims to learn a function $f: D \mapsto Y$.  

The input features of JIT-DP can be categorized into two \new{types~\cite{jitfine}}, namely, expert features and semantic features.
JIT-DP based on expert features~\cite{How_Well_Do_Change, expert_feature, CFeatures} hypothesizes that there is a fixed relationship between software metrics (expert features) and defects. These techniques define code change features as macro metrics such as lines of code added (LA), number of modified subsystems (NS), and developer experience (EXP) based on experts' understanding of how code commits cause \new{defects~\cite{jitfine}}. \new{They~\cite{How_Well_Do_Change, expert_feature, CFeatures}} use statistical machine learning models such as decision trees, support vector machines, and Bayesian networks to predict future defects.
These approaches have strong interpretability~\cite{jitfine, jitdp_Interpretability} while not relying on large datasets due to the experience of the experts. However, they often oversimplify the causes of software defects and overlook the intrinsic semantic factors.

In contrast, JIT-DP based on semantic features~\cite{cc2vec, deepjit, CodeJIT} identifies defect-prone changes by learning the semantic representations of code commits. \new{These approaches~\cite{cc2vec, deepjit, CodeJIT}} employ deep neural networks to encode code changes into discrete vector representations and acquires predictions through training a machine learning classifier. As a data-driven technology, it can learn code change semantics from large-scale code change corpora, thus reducing the labor required for feature extraction from human experts. However, deep learning approaches are \new{data-hungry~\cite{data_hungry,data_hungry_2}} at the same time, which hinders their effectiveness under the limited number of annotated samples in defect prediction datasets.

Inspired by the pros and cons of each feature type, most recent works~\cite{jitfine, SimCom} have explored the mixture of both, improving their approaches by combining the advantages of the two features.

\subsection{Code Change Representations}

Apart from JIT-DP, due to the swift advancement of deep learning, code representation learning has also emerged as a widespread technique~\cite{Commit2Vec}. This technology transforms code changes into discrete semantic vectors through deep neural networks~\cite{Represent_Edits}, particularly pre-trained language models (PLMs). 
Formally, let $x$=$c\oplus m$ denote a code commit fragment where $c$=\{$c_1$, $\ldots$, $c_{N_1}$\} is a change fragment with $N_1$ tokens, and $m$=\{$m_1$,$\ldots$, $m_{N_2}$\} represents the corresponding commit message with $N_2$ tokens. The goal of code change representation learning is to map $x$ to a $d$-dimensional vector that captures the semantics of the code change, i.e., $f_\theta:X\rightarrow\mathbb{R}^d$~\cite{Represent_Edits}, where $f_\theta$ is a learnable function parameterized by $\theta$. $f_\theta$ can be implemented using deep neural networks such as MLP~\cite{code2vec}, LSTM~\cite{LSTM}, and Transformer~\cite{Transformer}. 
The learned features are utilized for a broad range of downstream tasks such as change quality estimation~\cite{CodeReviewer}, automating code review~\cite{CodeReviewer}, and commit message generation~\cite{general_purpose_embeddings}.

Code change representation learning can be broadly categorized into two \new{types~\cite{cc2vec}}: sequence-based and graph-based approaches.
Sequence-based approaches treat code changes as token sequences and employ deep learning models, such as Convolutional Neural Networks (CNN)~\cite{deepjit}, Long Short-Term Memory Networks (LSTM)~\cite{How_Well_Do_Change}, and Transformers~\cite{CodeReviewer, CoreGen}, to learn the semantic representations. 

Graph-based approaches~\cite{Commit2Vec, Represent_Edits, jitast, structural_model, ccs2vec, CCcontext, COMU}, parse source code into abstract syntax trees (ASTs) and represent them as graphs. Nodes in the graph represent files or functions, while edges represent the dependencies and change relationships between \new{them~\cite{jitast}}. Structural representation enables the model to capture more complex relationships between code structures and \new{changes~\cite{jitast}}.

\begin{figure}[t]
\centerline{\includegraphics[width=\linewidth]{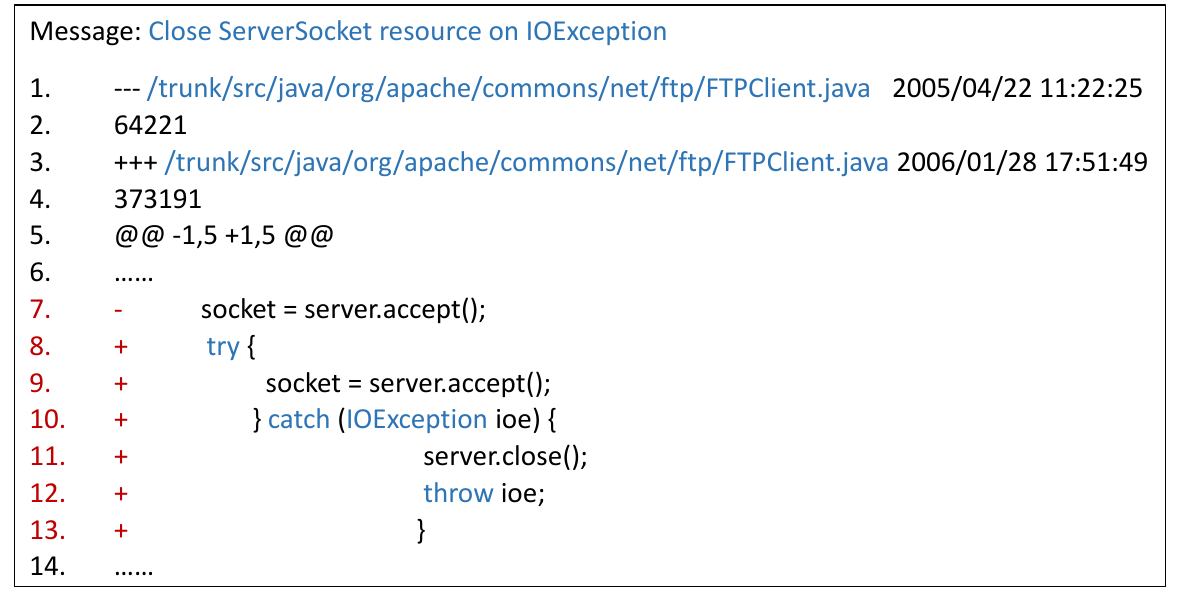}} 
\caption{An illustration of the importance of commit messages}
\label{example_commit_message}
\end{figure}

However, the limitation of previous research is that they do not fully utilize the semantic associations between commit messages and code changes. To illustrate how important commit messages can be for understanding code changes, an example is given in Figure \ref{example_commit_message}, namely, the commit \#373191 in Apache's project commons-net.

In this commit, developers wrapped "socket = server.accept();" (Line 7) with a try-catch block to close the server on an IOException by "server.close();" (Line 11). The logic in the above code changes is precisely described by its attached commit message: "Close ServerSocket resource on IOException." 

Inspired by the guidance that commit messages can help developers to understand code changes, \textbf{we intuitively hypothesize that code change representation models can learn more accurate code semantics guided by commit messages}. To verify our hypothesis and fill the gap in previous research, our work eventually designs a bi-modal change representation model incorporating both code change and commit message.


\section{Approach}


To fully exploit the semantics in commit messages, we propose a bi-modal change representation model for JIT-DP, called \model (Section 3.1), and then incorporate the model into the general framework of JIT-DP, denoted as \approach (Section 3.2). \model follows the ``pre-training \& fine-tuning'' paradigm of PLMs, wherein we pre-train a Transformer encoder on a large-scale change corpus and transfer the pre-trained knowledge to JIT-DP as its downstream task.

\subsection{Learning Code Change Representations}

\model extends \new{CodeBERT~\cite{codebert}} by incorporating commit messages into code change representations. 
Besides \new{masked language modeling (MLM)~\cite{bert}}, we design a new objective of pre-training, replaced message identification (RMI). By pre-training on these two objectives, the model performs bidirectional language modeling on code commits and explicitly captures the semantic correlations between commit messages and code changes. 

\subsubsection{Model Architecture}

\begin{figure}[!t]
\centerline{\includegraphics[width=1\linewidth,trim=30 0 30 10,clip]{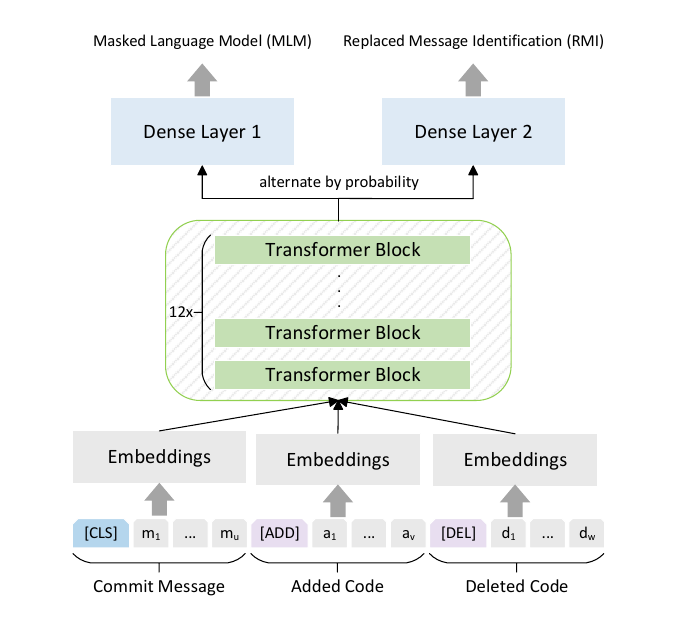}}
\caption{The architecture of \model}
\label{model_pretrain}
\end{figure}

Figure \ref{model_pretrain} illustrates the architecture of \model. Initially, the input code commit is tokenized and embedded through an embedding layer before feeding into a Transformer encoder to extract the change representation. The extracted change representation vectors are taken as input to two dense layers for pre-training the bi-modal change representation learning model.

To incorporate both commit message and code change into change representation learning, \model integrates three channels of information (i.e., commit message, added lines, and deleted lines) as input, separated with special tokens. A "[CLS]" token is added before the commit message, an "[ADD]" token is added before each "added line", and a "[DEL]" token is added before each "deleted line". This enables the model to differentiate different components from code commits during pre-training and the downstream defect prediction task. The commit message, added lines, and deleted lines are then tokenized and fed into \model to generate the corresponding code change representation.

\subsubsection{Masked Language Modeling (MLM)}

To enable \model to understand the deep semantics in code changes and commit messages, we firstly employ the widely used pre-training task \new{"Masked Language Modeling" (MLM)~\cite{bert}}. MLM replaces a small portion of tokens in the input code with a special [MASK] symbol and trains the \model to predict the original tokens given the masked code, thereby encouraging the model to utilize the contextual information for semantic representations.


Formally, given a text sequence $\boldsymbol{x} = {x_1, \ldots, x_n} \in X$, we randomly mask $m$ positions and create a masked sequence:


\begin{equation}
\begin{aligned}
\boldsymbol{x}^{(m)} & = x_1, \ldots, x_{i_1-1}, \mathtt{[MASK]},\\
& \ \ \ \ \ \ x_{i_1+1}, \ldots, x_{i_m-1}, \mathtt{[MASK]}, x_{i_m+1}, \ldots, x_n
\end{aligned}
\end{equation}

where $i_1, i_2, \ldots, i_m$ are randomly generated positions for masking. 


\new{The MLM task~\cite{bert}} aims to maximize the probability of predicting the correct words given the masked sequences. The loss function is defined in the form of negative log-likelihood:

\begin{equation}
\mathcal{L}_{mlm}(\boldsymbol{\theta})=\sum_{\boldsymbol{x}^{(m)} \in \mathcal{D}} \log p\left(x_{i_1}, x_{i_2}, \ldots, x_{i_m} \mid \boldsymbol{x}_{\neg m} ; \boldsymbol{\theta}\right)
\end{equation}
where $\mathcal{D}$ represents the training dataset, $\boldsymbol{x}_{\neg m}$ denotes all the tokens in $\boldsymbol{x}$ except for the masked positions, and $\boldsymbol{\theta}$ represents the model parameters. The objective of the negative log likelihood loss is to maximize the conditional probability $p(x_{i_1}, x_{i_2}, \ldots, x_{i_m} | \boldsymbol{x}_{\neg m}; \boldsymbol{\theta})$ to train the model parameters $\boldsymbol{\theta}$, enabling the model to predict the missing words correctly given the masked positions.

\subsubsection{Replaced Message Identification (RMI)}

The MLM pre-training task enables the \model to capture the semantic relationships between code tokens and the surrounding context in JIT-DP. However, it does not consider the bi-modal nature of change representation which involves both change patterns and the corresponding semantics in the commit messages. 

To explicitly learn the semantic relationship between commit messages and code changes, we propose a new pre-training objective named Replaced Message Identification (RMI). RMI randomly determines whether to replace the commit message with a non-corresponding message and trains the \model to predict whether the commit message is replaced, i.e., whether the commit message corresponds to its code change.

The pre-training objective of RMI is formulated as a binary classification task: given a code commit $x_i$, the model predicts whether the message has been replaced ($y_i=0$) or not ($y_i=1$). The prediction function $f_{RMI}$ can be learned by optimizing the cross-entropy loss function:



\begin{equation}
\begin{aligned}
\mathcal{L}_{rmi} (\theta) & =-\log \left(\prod_{i=1}^N p\left(y_i \mid x_i;\boldsymbol{\theta}\right)\right) \\
& =-\sum_{i=1}^N \left[\left(1-y_i\right) \log \left(1-p\left(y_i \mid x_i\right. 
;\boldsymbol{\theta})\right)\right. \\
& \left.+ y_i \log \left(p\left(y_i \mid x_i;\boldsymbol{\theta}\right)\right)\right]
\end{aligned}
\end{equation}
where $\boldsymbol{\theta}$ represents the model parameters, and $p(y_i|x_i;\boldsymbol{\theta})$ denotes the probability score from the model's output layer.

To accomplish this objective, the model needs to understand the code changes and the commit messages separately, and compare their consistency. Whereby, the model is forced to learn the explicit semantic relationship between commit messages and code changes, as well as utilize the key information provided by commit messages to better understand code changes in JIT-DP.

In the pre-training phase, \approach performs the MLM and RMI objectives alternately. This allows the model to learn shared features and knowledge from both MLM and RMI objectives.

\begin{figure}[!t]
\centerline{\includegraphics[width=\linewidth]{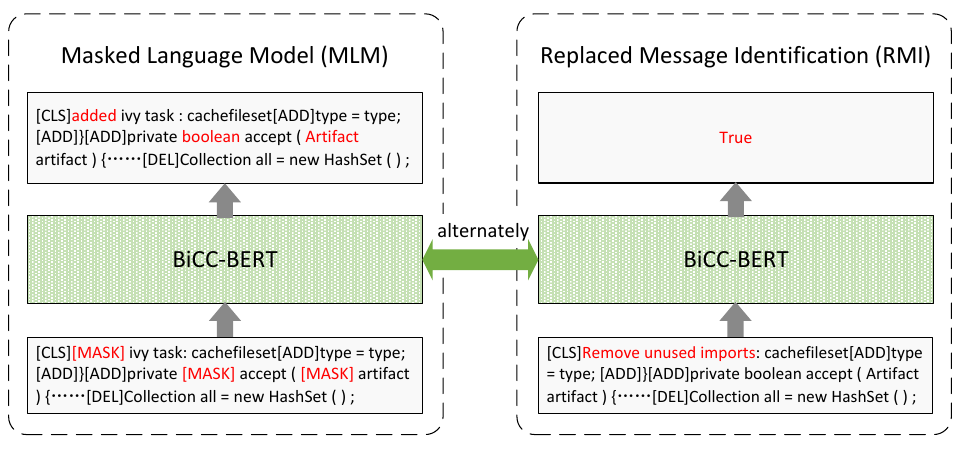}}
\caption{Illustration of the pre-training objectives}
\label{example}
\end{figure}

Figure \ref{example} illustrates an example of the MLM and RMI objectives. In the MLM example, a few tokens in the code commit are replaced with [MASK] symbols, and \model is trained to recover these masks. In the RMI example, the commit message is replaced by a random message irrelevant to the commit, and \model is trained to label this replaced message as ``True''.

\subsection{Just-In-Time Defect Prediction}

\begin{figure*}[!t]
\centerline{\includegraphics[width=0.95\linewidth,trim=0 5 0 5,clip]{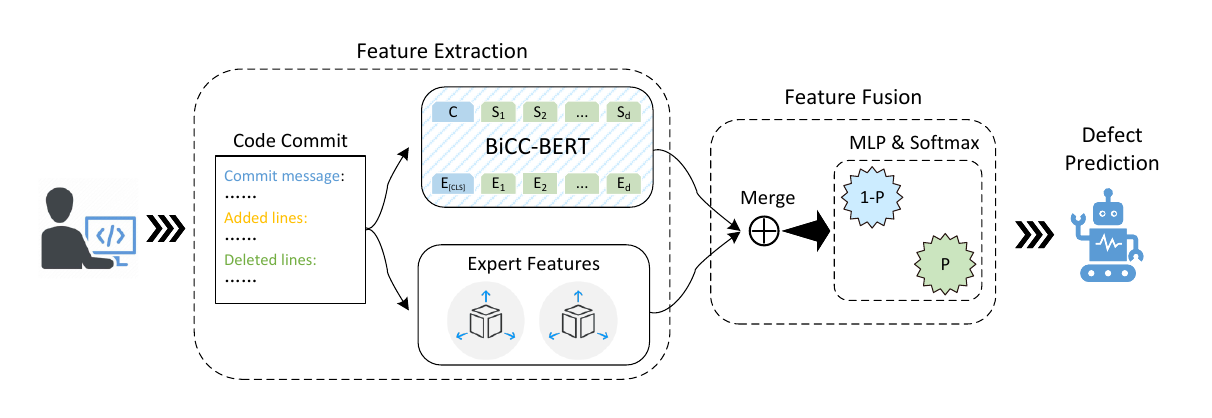}}
\caption{Overview of \approach}
\label{model_jitdp}
\end{figure*}

Based on the bi-modal change representation model, we propose a novel defect prediction approach called \approach by integrating \model into the general defect prediction framework. Figure \ref{model_jitdp} illustrates the overall architecture of \approach. Given code commits, \approach extracts two types of features, namely, code change representations learned by \model and 14 metric-based expert features. The extracted features and change representations are fused by concatenation. Finally, an MLP classifier is employed to predict defects based on the fused feature vector. 

The framework can be formulated as follows: given a code commit snippet $x_i$ and a code change representation model $g_\phi$, \approach first extracts the semantic feature vector $s_i = g_\phi(x_i)$ from the code commit snippet with the code change representation model. Then, \approach extracts the expert feature vector $e_i = h(x_i)$ using a function $h$ predefined by experts. The fused feature vector of the semantic and expert features is denoted as $m_i = s_i \oplus e_i$. Subsequently, the fused feature vector is fed into the JIT-DP model $f_{DP}$, which predicts the probability of the code commit snippet containing defects as $p(y_i|x_i) = f_{DP}(m_i)$, where $y_i$ indicates whether the code commit snippet contains a defect.

\subsubsection{Feature Extraction and Fusion}

To fully leverage the code commit representations that \model learns from pre-training, \approach directly inputs code commits into the \model to extract semantic features from code commits.

At the same time, \approach extracts expert features based on predefined metrics. \approach represents expert features as a 14-dimensional vector, each dimensionality corresponding to a widely used change-level code metric proposed by Kamei et al.~\cite{expert_feature}. The list of all 14 metrics is presented in Table \ref{expert_features}. 

To make the most utility of both expert and semantic features, \approach subsequently fuses them into one vector and takes them as input to a classifier. As the change representation vectors have a different dimensionality to that of the expert feature vector, \approach follows previous works~\cite{jitfine, deeper} and expands the expert features to the same dimensionality of the change representation vector through a trainable dense layer. Finally, \approach concatenates the two vectors into a new vector, which serves as the fused feature vector for the entire code commit.

\subsubsection{Defect Prediction}

The defect prediction model takes the fused vector of semantic and expert features as input and predicts the probability of the corresponding code commit being defective. The model maps the fused vector to a 2-dimension vector using a multi-layer perceptron (MLP) and a softmax layer. The 2-dimension vector consists of two elements respectively representing the probabilities of the code commit being defective and defect-free. Finally, the model's parameters are optimized by minimizing the loss function, which measures the difference between these probabilities and the true labels. During the fine-tuning of JIT-DP, the parameters of the code change representation model and the defect prediction model are trained simultaneously.

The JIT-DP model can be learned by minimizing the cross-entropy loss function:



\begin{equation}
\begin{aligned}
\mathcal{L} & =-\log \left(\prod_{i=1}^{|\mathcal{D}|} p\left(y_i \mid m_i;\boldsymbol{\theta}\right)\right) \\
& =-\sum_{i=1}^{|\mathcal{D}|} \left[\left(1-y_i\right) \log \left(1-p\left(y_i \mid m_i;\boldsymbol{\theta}\right)\right)\right. \\
& \left.+ y_i \log \left(p\left(y_i \mid m_i;\boldsymbol{\theta}\right)\right)\right]
\end{aligned}
\end{equation}
where $\boldsymbol{\theta}$ represents the model parameters, and $p(y_i|m_i;\boldsymbol{\theta})$ denotes the probability score from the model's output layer.

\begin{table*}[htbp]
\caption{The 14 expert features used in our study}
\begin{center}
\begin{tabular}{cccc}
\toprule
\textbf{Index} & \textbf{Name} & \textbf{Description} & \textbf{Dimension} \\
\midrule
1 & NS & Number of modified subsystems & Diffusion\\
2 & ND & Number of modified directories & Diffusion\\
3 & NF & Number of modified files & Diffusion\\
4 & Entropy & Distribution of modified code across files & Diffusion\\
5 & LA & Number of lines added & Size\\
6 & LD & Number of lines deleted & Size\\
7 & LT & Number of lines in the file before modification & Size\\
8 & FIX & Whether the change is a defect fix & Purpose\\
9 & NDEV & Number of developers who changed the files & History\\
10 & AGE & Average time interval between the latest historical records & History\\
11 & NUC & Number of unique changes made to the modified files & History\\
12 & EXP & Developer experience & Experience\\
13 & REXP & Recent developer experience & Experience\\
14 & SEXP & Developer experience on the subsystem & Experience\\\bottomrule
\end{tabular}
\end{center}
\label{expert_features}
\end{table*}

\section{Experimental Setup}

\subsection{Research Questions}

To investigate the effectiveness of \approach, we aim to answer the following three research questions:

\emph{RQ1. How does \approach perform in JIT-DP?}
We compare \approach with state-of-the-art approaches from different categories on the same dataset to evaluate the effectiveness of \approach.

\emph{RQ2. How effective is the bi-modal change representation approach?}
We ablate the MLM, the RMI, and both objectives individually from \approach and conduct comparative experiments to evaluate the effectiveness of the bi-modal change representation.

\emph{RQ3. How can the combination of MLM and RMI affect the performance of \approach?}
We explore the impact of different combinations of the two objectives on the performance of \approach, including their training orders and ratios of sampling probability. By conducting comparative experiments, we aim to determine the most favorable combination of pre-training objectives in JIT-DP.

\subsection{Dataset}

We train and test \approach on JIT-Defects4J~\cite{jitfine}, a commonly used benchmark for JIT-DP. \new{JIT-Defects4J is derived from the manually \new{labeled} LLTC4J (Line-Labelled Tangled Commits for Java) dataset~\cite{LLTC4J}, with thorough preprocessing (e.g., ignoring excessively large commits, excluding commits with no added code lines, and filtering out non-functional code changes) to ensure high quality. In contrast, traditional JIT-DP datasets such as Hoang et al.'s~\cite{cc2vec,deepjit} and Zeng et al.'s~\cite{how_far_are_we}, are automatically annotated using the SZZ algorithm~\cite{SZZ}, which was shown to introduce a large number of false positives and negatives~\cite{ccs2vec,dataset_worry1,dataset_worry2,dataset_worry3} due to tangled commits. JIT-Defects4J has also been widely used in recent studies on JIT-DP such as JIT-Fine~\cite{jitfine} and CCT5~\cite{CCT5}. To establish a fair comparison, we use the same training and testing datasets as recent works on JIT-DP~\cite{jitfine, Dataset_multi-objective, CCT5, Dataset_PEFT}. }The dataset consists of 27,319 code commit records from 21 Java open-source projects (e.g., opennlp, commons-io, and commons-math), including 2,332 defective ones and 24,987 non-defective ones, enabling the demonstration of the practical application of \approach on real-world projects.

\subsection{Evaluation Metrics}

We employ two commonly used metrics, F1-score and AUC, to evaluate the performance of JIT-DP. 

\textbf{F1-score:} a metric for measuring the overall accuracy of a JIT-DP model. It is calculated as the harmonic mean of precision and recall and ranges between 0 and 1, with 1 being the best score. F1-score can be computed as: $$F1=\frac{2\times Recall\times Precision}{Recall+Precision}$$ where precision is the ratio of true defective predictions to the total number of defective predictions, and recall is the ratio of true defective predictions to the total number of actual defective instances.

\textbf{AUC (Area Under the ROC Curve):} a widely used metric for measuring the model's ability to discriminate between defective and non-defective commits across different threshold settings. The ROC curve is created by plotting the true positive rate (sensitivity) against the false positive rate (1 - specificity) at various classification thresholds. The AUC ranges between 0 and 1, with 1 indicating a perfect classifier.

\subsection{Comparison Approaches}

We compare \approach with 8 state-of-the-art approaches, including LApredict~\cite{how_far_are_we}, Yan et al.'s work~\cite{Yanetal}, Deeper~\cite{deeper}, DeepJIT~\cite{deepjit}, JITLine~\cite{JITLine}, CC2Vec~\cite{cc2vec}, JIT-Fine~\cite{jitfine}, and CodeReviewer~\cite{CodeReviewer}. 
Table~\ref{baseline_intro} gives a comparative summary of these approaches on various components, including code change representation models, \new{considered features for defect prediction}, and the use of commit messages. 




\begin{table*}[htbp]
\caption{A Comparative Summary of Baselines and \approach}
\label{baseline_intro}
\begin{center}
\begin{threeparttable}
\begin{tabular}{cccc}
\toprule
\textbf{Approach} & \textbf{Code Change Representation Model} & \new{\textbf{Features for Defect Prediction}} & \textbf{Use of Commit Messages}\\\midrule
LApredict~\cite{how_far_are_we} & None & \new{Expert Features} & None \\
Yan et al.~\cite{Yanetal} & None & \new{Token and Expert Features} & None \\
Deeper~\cite{deeper} & None & \new{Expert Features} & None \\
DeepJIT~\cite{deepjit} & None & \new{Semantic Features} & Implicitly Learning Associations \\
JITLine~\cite{JITLine} & None & \new{Token and Expert Features} & None \\
CC2Vec~\cite{cc2vec} & Hierarchical Attention Network & \new{Semantic Features} & Implicitly Learning Associations \\
JIT-Fine~\cite{jitfine} & Code Pre-training Model & \new{Semantic and Expert Features} & Implicitly Learning Associations \\
CodeReviewer~\cite{CodeReviewer} & Code Change Pre-training Model & \new{Semantic Features} & None \\
\approach & Bi-modal Change Pre-training Model & \new{Semantic and Expert Features} & Explicitly Learning Associations \\\bottomrule

\end{tabular}

\end{threeparttable}
\end{center}
\end{table*}

\begin{itemize}
\item[1)] \textbf{LApredict}~\cite{how_far_are_we}: a logistic regression classifier using the feature "added code lines". 

\item[2)] \textbf{Yan et al.'s work}~\cite{Yanetal}: a defect prediction model using logistic regression on expert features, accompanied with a defect localization model using N-grams. 

\item[3)] \textbf{Deeper}~\cite{deeper}: a defect prediction model using deep belief networks. 

\item[4)] \textbf{DeepJIT}~\cite{deepjit}: a defect prediction model by learning semantic features of code commits using convolutional neural networks. 

\item[5)] \textbf{JITLine}~\cite{JITLine}: a defect prediction model using both expert and token features with a random forest classifier, accompanied by a defect localization model based on token features using local interpretable model-agnostic explanations (LIME) \cite{LIME}. 

\item[6)] \textbf{CC2Vec}~\cite{cc2vec}: a defect prediction model that learns code change representations based on a hierarchical attention network. 

\item[7)] \textbf{JIT-Fine}~\cite{jitfine}: a unified model for defect prediction and localization through a combination of semantic and expert features. 

\item[8)] \textbf{CodeReviewer}~\cite{CodeReviewer}: a pre-trained code change representation model. Unlike \approach, CodeReviewer only focuses on code review tasks such as code change quality assessment. For a fair comparison, we fine-tune and evaluate CodeReviewer on our JIT-DP dataset.

\end{itemize}

\new{The baselines we select encompass a variety of types. LApredict, Yan et al.’s work, Deeper, DeepJIT, and JITLine are not based on code change representation learning. They are compared to assess the effectiveness of using code change representation models. CC2Vec utilizes a hierarchical attention network without pre-training. We choose it to assess the impact of pre-training. Similarly, JIT-Fine is included to compare the effectiveness between code and change representations. Finally, CodeReviewer is compared to evaluate the effectiveness of the bi-modal change representation learning.}
\new{We also consider the variety of features for defect prediction when we select baselines. Our baselines encompass all features used in recent studies of JIT-DP, including expert, token, and semantic features. This allows us to comprehensively compare the performance of our approach with those based on all categories of JIT-DP features.}


\new{A more detailed literature review of existing approaches will be provided in Section 6.}

\subsection{Implementation Details}

We build \model based on RoBERTa~\cite{RoBERTa} using the default configuration as RoBERTa-base (H=768, A=12, L=12). The parameters of the \model are initialized with the checkpoint \texttt{microsoft/codebert-base}~\cite{codebert}. We also use the default CodeBERT tokenizer. Following JIT-Fine~\cite{jitfine}, we add two special tokens, "[ADD]" and "[DEL]", to the original tokenizer's vocabulary (with a size of 50,265). The maximum input sequence length is set to 512. We implement all models based on the \new{Huggingface Transformers~\cite{huggingface}}. All models are trained using Adam~\cite{Adam} with FP16 on two Nvidia GeForce RTX 3090 GPUs. 

During the pre-training phase, we set the batch size per GPU to 16 and the gradient accumulation steps to 32. The number of training epochs is set to 100. The initial learning rate is set to 5e-4 and linearly increases from 0 during the warm-up phase. The masked token proportion in the MLM objective is set to 15\%. For each training batch, the \model is alternately trained using one of the MLM and RMI objectives (with a probability ratio of 2:1).

During the training phase of JIT-DP, the batch size per GPU is set to 12, and the initial learning rate is set to 1e-5. Similar to the pre-training phase, the learning rate of the JIT-DP model linearly increases from 0 during the warm-up phase. The performance on the validation set is evaluated during the training process, and the checkpoint of the model with the best F1 score on the validation set is selected for testing.

\section{Experimental Results}

\subsection{Effectiveness of \approach (RQ1)}



\begin{table}[!t]
\caption{Performance comparison across different JIT-DP approaches. The highest scores are highlighted in bold (similarly in Tables \ref{table_ablation_on_objective}, \ref{table_ablation_on_shuffle}, and \ref{table_ablation_on_probability}). }
\label{table_compare_to_baseline}
\begin{center}
\begin{tabular}{ccc}
\toprule
\textbf{Approach} & \textbf{F1-score} & \textbf{AUC} \\\midrule
LApredict & 0.059 & 0.694 \\
Work of Yan et al. & 0.062 & 0.675 \\
Deeper & 0.246 & 0.682 \\
DeepJIT & 0.293 & 0.775 \\
JITLine & 0.261 & 0.248 \\
CC2Vec & 0.248 & 0.791 \\
JIT-Fine & 0.431 & 0.881 \\
CodeReviewer & 0.341 & 0.802 \\
\approach & \textbf{0.478} & \textbf{0.887} \\\bottomrule

\end{tabular}
\end{center}
\end{table}

\new{We evaluate the overall performance by comparing \approach with baseline models outlined in Section 4. }
The results are presented in Table \ref{table_compare_to_baseline}. We observe that \approach outperforms all other comparative approaches in all evaluation metrics. \new{Compared to JIT-Fine, the highest-performing baseline, \approach achieves a relative performance improvement of 10.8\% in terms of F1-score and 0.7\% in terms of AUC. }\new{The significant improvement in F1-score balances recall and precision for JIT-DP, enabling software developers to better identify potential bugs in their code commits with minimal human effort based on \approach’s predictions.}


We also find that JIT-Fine, CodeReviewer, and \approach perform noticeably better compared to other approaches in terms of both F1-score and AUC. These approaches are the only ones that build change representation models based on pre-training approaches. This implies that pre-trained change representation models can contribute to a deeper understanding of code commits. Furthermore, JIT-Fine and \approach achieve the best results among all baselines, being the only two approaches that take advantage of both expert features and semantic features. This indicates that feature fusion can also benefit the performance of JIT-DP, \new{which is one of the reasons that \approach outperforms other approaches.}

Another interesting point is that CodeReviewer also achieves competitive performance compared to all comparative approaches except JIT-Fine and \approach. This is probably because CodeReviewer is also pre-trained to learn the code change representations, which contributes to more accurately identifying defective code commits. However, as this model is primarily trained to automate code review tasks, it fails to outperform JIT-Fine and \approach, which are designed to focus on JIT-DP. \new{This highlights the importance of constructing a specialized code change representation learning model for JIT-DP, which is essential for \approach to outperform others.}

\smallskip\textbf{Answer to RQ1:}
\new{\approach achieves state-of-the-art performance on JIT-DP, significantly aiding software developers in prioritizing their efforts to hunt for potential code defects. Our bi-modal change pre-training model (\model) facilitates a more effective understanding of code changes.}

\subsection{Effectiveness of Bi-modal Change Representation (RQ2)}


\begin{table}[htbp]
\caption{Impact of pre-training objectives on JIT-DP.}
\label{table_ablation_on_objective}
\begin{center}
\begin{tabular}{lcc}
\toprule
\textbf{Approach} & \textbf{F1-score} & \textbf{AUC} \\
\midrule
\approach & \textbf{0.478} & \textbf{0.887} \\
 \ -w/o RMI \& MLM & 0.431 & 0.881 \\
 \ -w/o RMI & 0.452 & 0.884 \\
 \ -w/o MLM & 0.397 & 0.854 \\
\bottomrule
\end{tabular}
\end{center}
\end{table}

Table \ref{table_ablation_on_objective} presents the results of ablation studies. \new{
We compare \approach with three variants of code change representation models: \model without the RMI and MLM objectives, \model without the MLM objective, and \model without the RMI objective. For the variants that require pre-training, we maintain the same quantity of data samples and number of training epochs as used in \approach.} We find that \approach outperforms all variants in terms of F1-score and AUC, indicating that both MLM and RMI are effective for bi-modal change representation learning. 
This implies that the contextual modeling approach can enhance the model's understanding of the entire code commit by allowing the model to gradually learn the dependencies between different parts of a code commit. Additionally, explicitly capturing the semantic correlation between commit messages and code changes can further enhance the model's understanding of code changes. Therefore, both the RMI and MLM objectives contribute to strengthening the \model's understanding of code commits and achieving better performance in JIT-DP.

We also observe that pre-training \approach with only the RMI objective (i.e., \approach without the MLM objective) performs even worse compared to no pre-training (i.e., \approach without RMI and MLM objectives). This is probably because the parameters of \model are initialized with the checkpoint of \texttt{microsoft/codebert-base}~\cite{codebert}, which is only trained on a code corpus instead of a code-change corpus. \new{This implies that adapting \model to the code-change corpus through continuous pre-training with the MLM objective allows the model to benefit from the RMI objective, leading to performance improvements.}

\smallskip\textbf{Answer to RQ2:}
\new{Bi-modal change representation learning plays a crucial role in JIT-DP, enabling our proposed \approach to grasp deeper semantics in code commits. Pre-training with both MLM and RMI enhances the model's effectiveness in JIT-DP.}

\subsection{Impact of Pre-training Objective Combinations (RQ3)}

We further investigate the impact of different combinations of pre-training objectives, including training orders and sampling probability on JIT-DP. 

\subsubsection{Training Order of Pre-training Objectives}

\new{
We examine the standard \approach alongside two configurations for training the bi-modal change representation model: 1) pre-training with RMI followed by MLM (referred to as RMI$\rightarrow$MLM), and 2) pre-training with MLM followed by RMI (referred to as MLM$\rightarrow$RMI). 
}

\begin{table}[htbp]
\caption{JIT-DP performance under different orders of pre-training objectives}
\label{table_ablation_on_shuffle}
\begin{center}
\begin{tabular}{ccc}
\toprule
\textbf{Pre-training Order} & \textbf{F1-score} & \textbf{AUC} \\\midrule
RMI$\rightarrow$MLM & 0.417 & 0.885 \\
MLM$\rightarrow$RMI & 0.456 & 0.883 \\
Alternating RMI \& MLM & \textbf{0.478} & \textbf{0.887} \\\bottomrule
\end{tabular}
\end{center}
\end{table}

The results in Table \ref{table_ablation_on_shuffle} show the impact of different training orders of pre-training objectives on the performance of JIT-DP. The default setting of \approach outperforms all the variant settings in terms of F1-score and AUC, indicating that alternating the training of MLM and RMI significantly improves the evaluation metrics compared to training MLM and RMI successively. 
Therefore, it can be concluded that training \approach by alternating MLM and RMI objectives with probability-based sampling plays a crucial role in improving its performance. The reason is that training MLM and RMI objectives successively leads to catastrophic forgetting, where the model forgets the knowledge learned from the previous objective while training the latter one. Furthermore, alternating the training of MLM and RMI objectives facilitates mutual supervision between these two pre-training objectives, allowing the model to share and complement objective-relevant knowledge. Thus, alternately training MLM and RMI objectives can help \model further enhance the understanding of code commits and achieve better generalization ability, improving the performance of JIT-DP.

An interesting point is that pre-training the change representation model with RMI followed by MLM (i.e., RMI$\rightarrow$MLM) does not contribute to performance improvement. The underlying reason is perhaps the same as why pre-training \approach with only the RMI objective performs even worse compared to no pre-training, as we have discussed in RQ2. 
\new{These results underscore the importance of domain adaptation for achieving improved performance.}

\subsubsection{Objective Proportion}
We contrast the default setting of \approach (referred to as 2$\times$MLM, 1$\times$RMI) with two variants in the bi-modal change representation model. These variants involve different ratios for combining training objectives: 1) pre-training with MLM and RMI with 1:1 ratio (referred to as 1$\times$MLM, 1$\times$RMI), and 2) pre-training with MLM and RMI objectives with 3:1 ratio (referred to as 3$\times$MLM, 1$\times$RMI). 


\begin{table}[htbp]
\caption{JIT-DP performance under different probability ratios for selecting pre-training objectives}
\label{table_ablation_on_probability}
\begin{center}
\begin{tabular}{ccc}
\toprule
\textbf{Ratios} & \textbf{F1-score} & \textbf{AUC} \\\midrule
1$\times$MLM, 1$\times$RMI & 0.436 & 0.880 \\
2$\times$MLM, 1$\times$RMI & \textbf{0.478} & \textbf{0.887} \\
3$\times$MLM, 1$\times$RMI & 0.455 & 0.884 \\\bottomrule
\end{tabular}
\end{center}
\end{table}

The results in Table \ref{table_ablation_on_probability} show that the sampling ratio of the pre-training objectives has a significant impact on the performance of \approach. Among all evaluation metrics, pre-training \approach with a sampling ratio of 2:1 outperforms all other variants. 
The reason is that there is a mutually supervisory relationship between the MLM and RMI training objectives, and an appropriate sampling ratio allows the model to better complement the knowledge learned from alternating training. On the other hand, excessively high or low ratios hinder the formation of a good mutual supervisory relationship between the two pre-training objectives. Hence, combining the MLM and RMI objectives with a sampling ratio of 2:1 best benefits \model in extracting deeper semantic information from code commits, thereby exhibiting better performance in JIT-DP.

\smallskip\textbf{Answer to RQ3:}
\new{Alternately training \model with MLM and RMI using an appropriate ratio (2:1)} fosters mutual supervision between the objectives, thereby enhancing the model's understanding of code commits and leading to optimal performance in JIT-DP. 

\subsection{Threats to Validity}



\smallskip\textbf{Threats to external validity} primarily pertain to the characteristics of the dataset under study. Due to the limitation of our dataset, JIT-Defeat4J, the diversity of projects considered in our research is limited in three key aspects.

First, our analysis focuses solely on projects developed in Java, overlooking other popular languages such as C/C++ and Python. Incorporating projects from these languages would provide a more comprehensive perspective.

Second, the projects analyzed in our study are exclusively open-source projects, and we have no knowledge of \approach's performance on commercial projects. 

Lastly, we must acknowledge that not all software bugs can be identified within the code repository. Despite the manual labeling conducted in LLTC4J~\cite{LLTC4J}, there is still a possibility of missing certain bug-fixing commits.

\smallskip\textbf{Threats to construct validity} correspond primarily to the selection of evaluation metrics in our experiments. To mitigate these threats, we have employed two different metrics, F1-score and AUC, which are widely used in previous studies. By employing these  metrics, we aim to provide a comprehensive and well-rounded assessment of our model's performance.

\section{Related Work}

\subsection{Code Change Representation Learning}
Code change representation learning, which aims to represent code changes as distributed vectors~\cite{Represent_Edits}, has attracted extensive attention of researchers in recent years. 
Compared to code representation, code change representation focuses more on the evolution of code repositories, allowing for the learning of the trajectory and patterns of code changes from the edit history in code repositories, and can therefore better capture fine-grained features of code changes and predict the trends of code evolution.

\smallskip\textbf{Sequence-based approaches.}
Researchers have proposed various sequence-based approaches for code change representation learning. Hoang et al.~\cite{cc2vec} proposed CC2Vec, which learns the representations of code changes based on a hierarchical attention network guided by commit messages. Nie et al.~\cite{CoreGen} introduced CoreGen, which utilizes the contextual information to automate the generation of commit messages  based on the Transformer model. Panthaplackel et al.~\cite{copy_that} extended the seq2seq model to copy an entire span from the input commit in a single step, reducing the number of model decisions required during inference. Pravilov et al.~\cite{general_purpose_embeddings} presented a novel approach to construct embeddings for code changes by unsupervised pre-training on a large unlabeled code change corpus. The embeddings are applied to the tasks of commit message generation and applying changes to code. Li et al.~\cite{CodeReviewer} introduced CodeReviewer, a pre-training model that utilizes four objectives to learn distributed representations for code changes and code reviews. Zhang et al.~\cite{CoditT5} proposed CoditT5, which learns representations for code changes through pre-training on a large dataset of source code and natural language comments. Zhou et al.~\cite{CCBERT} introduced CCBERT, a code change PLM that considers fine-grained information without relying on commit message. Lin et al.~\cite{CCT5} proposed CCT5, a generative code change PLM pre-trained on their own large-scale code change dataset. By pre-training on code change dataset, CCBERT and CCT5 can both learn better domain knowledge related to code change. 

\smallskip\textbf{Graph-based approaches.}
Graph-based approaches parse code into abstract syntax trees (ASTs) and represent them as graphs. Yin et al.~\cite{Represent_Edits} introduced a graph neural network-based approach that learns representations for code changes based on the AST of the code via an autoencoder framework. Brody et al.~\cite{structural_model} proposed a structured model for context code changes. Their model uses paths in the code change AST to represent the edits that occur in specific contexts. Lozoya et al.~\cite{Commit2Vec} presented commit2Vec, an AST-based approach that fully leverages the syntactic structure of code to automatically extract generic representations for code changes. Yao et al.~\cite{Incremental_Tree_Transformations} proposed a general model for incremental editing of code change ASTs that learns iterative edits on the trees, such as deleting or adding subtrees. Zhuang et al.~\cite{jitast} proposed ACE, an approach based on embeddings of AST changes, which compares the ASTs before and after code changes, capturing the associations in syntax structure. Zhang et al.~\cite{ccs2vec} proposed ccs2vec, which slices the code change program dependency graph for code changes and feeds the slices into a pre-trained sparse Transformer model. Vu et al.~\cite{CCcontext} presented a graph-based approach that encodes changes in the context of their surrounding code by converting the code changes into graphs, allowing the model to learn from the hierarchical and dependency structure inherent in code. Wang et al.~\cite{COMU} proposed COMU, a multi-grained contextual change representation learning approach that extracts information from the changed code at both the line and AST levels, enabling the model to understand code changes from different perspectives.

Compared to existing models, our model \model explicitly learns the semantic correlation between code changes and their corresponding commit messages. RMI pre-training objective enables the model to utilize the semantics in commit messages, thus leading to a better understanding of code changes.

\subsection{Just-in-Time Defect Prediction}

JIT-DP approaches identify changes that are likely to introduce defects based on historical changes. Existing approaches can be categorized into two types~\cite{jitfine}: expert-feature-based JIT-DP and semantic-feature-based JIT-DP.

\smallskip\textbf{Expert-feature-based JIT-DP.}
It utilizes predefined metrics by experts based on their understanding and capturing of defect factors in code commits~\cite{jitfine,How_Well_Do_Change,expert_feature}. 
Wen et al.~\cite{How_Well_Do_Change} proposed FENCES, which extracts six features covering different aspects of code changes and treats JIT-DP into a sequence labeling problem that is solvable by a recurrent neural network. Pornprasit et al.~\cite{JITLine} introduced JITLine, which extracts bag-of-words features from code changes and predict defective commits based on random forest. These expert-based approaches have strong interpretability~\cite{jitfine,jitdp_Interpretability} and do not require training models on extremely large datasets. However, expert-feature-based approaches oversimplify the causes of software defects and overlook the inherent semantic factors that trigger software defects.

\smallskip\textbf{Semantic-feature-based JIT-DP.}
Different from expert-feature-based JIT-DP, semantic feature-based JIT-DP learns semantic representations from commit messages and code changes and utilizes these representations to identify defect-prone changes. Hoang et al.~\cite{deepjit} proposed DeepJIT, a deep learning model that evaluates the likelihood of defects in code changes based on extracted semantic features. Following DeepJIT, they further proposed CC2Vec~\cite{cc2vec}. CC2Vec models the hierarchical structure of code changes using attention mechanisms and learns representations guided by the commit messages semantics, achieving superior performance compared to DeepJIT. Ni et al.~\cite{jitfine} introduced JIT-Fine, which establishes a unified model for both defect prediction and localization based on CodeBERT, while combining both semantic and expert features. Zhuang et al.~\cite{jitast} presented a gated hierarchical model called GH-ACE, which incorporates AST-based code change embeddings into JIT-DP, enabling the model to better capture the associations in code syntactics. These approaches benefit from learning semantic knowledge in large-scale code change corpora, thereby reducing the effort of manual feature extraction by human experts. Zhou et al.~\cite{SimCom} proposed SimCom++, a framework that combines expert-feature-based and semantic-feature-based models for JIT-DP, leveraging the complementary strengths of both two features.

Compared to existing approaches, \approach leverages bi-modal change representation learning, which enables it to capture rich self-supervised knowledge in commit messages semantics and thus achieves better performance in just-in-time defect prediction.

\section{Conclusion}

In this paper, we propose a bi-modal change representation model, named \model, for just-in-time defect prediction. We begin by designing two pre-training objectives to enhance \model's understanding of code changes under the guidance of the commit message. Based on \model, we introduce \approach, a novel JIT defect prediction approach by integrating the learned bi-modal change representations. Experimental results show that our model achieves a 10.8\% performance improvement in terms of F1-score compared to the state-of-the-art approaches. The results imply that commit messages play a critical role in guiding machine learning models to understand code change semantics. In the future, we will explore pre-training \model on larger bi-modal change corpora.

Our source code and experimental data for replication are publicly available at: \href{https://github.com/jyz-1201/JIT-BiCC/}{https://github.com/jyz-1201/JIT-BiCC/}. 

\section*{Acknowledgments}
This research is supported by National Key R\&D Program of China (Grant No. 2023YFB4503802) and  National Natural Science Foundation of China (Grant No. 62232003, 62102244 and 62032004).




\bibliographystyle{elsarticle-num}

\balance
\bibliography{ref}

\vskip400pt

\bio{}
Yuze Jiang received a bachelor’s degree from the School of Software at Shanghai Jiao Tong University, China, in 2023. He is currently a Ph.D. student in the Department of Computer Science and Engineering at the Ohio State University, United States. His research interests include intelligent software engineering and large language models.
\endbio

\bio{}
Beijun Shen is an associate professor in the School of Software, Shanghai Jiao Tong University. She received a Ph.D. degree from Institute of Software, Chinese Academy of Sciences, in 2001. Her research interests include LLMs for code and code intelligence. She has published over 200 papers in top conferences and journals in the field of software engineering, and received ICSE Distinguished Paper Award in 2023.
\endbio

\bio{}
Xiaodong Gu is currently an associate professor in the School of Software, Shanghai Jiao Tong University. He received a Ph.D. degree from the Department of Computer Science and Engineering from The Hong Kong University of Science and Technology, in 2017. His research interests lie in the broad areas of software engineering and deep learning, including large language models (LLMs) for code, code generation, code search, and code translation.
\endbio

\end{sloppypar}
\end{document}